\def\beq#1{\begin{equation}\label{#1}}
\def\eeq{\end{equation}}
\def\beqa#1{\begin{eqnarray}\label{#1}}
\def\eeqa{\end{eqnarray}}
\def\eq#1{equation~(\ref{#1})}
\def\Eq#1{Equation~(\ref{#1})}

\def\bfig{\begin{figure}[h] \centerline{\hbox{}}\vfill}
\def\efig{\end{figure}\vfill\newpage}

\def\fig#1{Figure~\ref{#1}}
\def\Fig#1{Figure~\ref{#1}}
\def\fignum#1{~\ref{#1}}

\def\spose#1{\hbox to 0pt{#1\hss}}
\def\simlt{\mathrel{\spose{\lower 3pt\hbox{$\mathchar"218$}}
     \raise 2.0pt\hbox{$\mathchar"13C$}}}
\def\simgt{\mathrel{\spose{\lower 3pt\hbox{$\mathchar"218$}}
     \raise 2.0pt\hbox{$\mathchar"13E$}}}
\def\simpropto{\mathrel{\spose{\lower 3pt\hbox{$\mathchar"218$}}
     \raise 2.0pt\hbox{$\propto$}}}

\def\addr#1{{\footnotesize\it #1}}

\def\pp{\noindent\parshape 2 0truecm 17truecm 2truecm 15truecm}
\def\rf#1;#2;#3;#4 {\par\pp#1, #2, #3, #4. \par}
\def\rn{\pp}
 
\def\etal{{\frenchspacing\it et al.}}
\def\ie{{\frenchspacing\it i.e.}}
\def\eg{{\frenchspacing\it e.g.}}

\def\tento#1{\times 10^{#1}}
\def\aet#1#2{\approx #1 \tento{#2}}

\def\Ms{M_{\odot}}     

\def\K{{\rm K}}
\def\s{{\rm s}}
\def\ergs{{\rm erg}}

\def\cm{{\rm cm}}
\def\eV{{\rm eV}}

\def\years{{\,\rm years}}

\def\x{x}
\def\f{f}
\def\nc{n_{cr}}
\def\Lrlte{L_r^{(lte)}}
\def\Lrlow{L_r^{(n\to 0)}}
\def\tc{\tau_{cool}}
\def\th{\tau_{h}}
\def\tg{\tau_{g}}

\def\recrate{k_1}
\def\molrate{k_2}
\def\goodrate{k_3}
\def\badrate{k_4}
\def\molratetwo{k_5}
\def\goodratetwo{k_6}
\def\badratetwo{k_7}
\def\comprate{k_8}

\def\HH{{\Lambda}}

\def\molcoolrate{\HH_m}
\def\linecoolrate{\HH_l}

\def\Lrec{\recrate}
\def\Lmol{\molrate}
\def\Lmoltot{k_m}

\def\Tvir{T_{vir}}
\def\zvir{z_{vir}}

\def\Oz{\Omega_0}
\def\Lz{\lambda_0}

\def\vr{{\bf r}}
\def\H{{\rm H}}

\def\Tg{T_\gamma}
\def\rhovir{\rho_{\rm vir}}

\def\ignore#1{}
\def\ed{\end{document}}

\documentstyle[11pt,rotate,epsf]{article}
\begin{document}

%%%%%%%%%%%%%%%%%%%%%%%%%%%%%

\begin{titlepage}   % Not numbered.

\noindent
\centerline{\hbox{\,}}
%\today
%\hfill MPI-PhT/95-666
%\vskip0.4truecm
\begin{center}
\vskip-2cm
{\bf
HOW SMALL WERE THE FIRST COSMOLOGICAL OBJECTS?
\footnote{Published in {\it ApJ}, {\bf 474}, 1-12 (January 1, 1997). Available from\\ 
{\it h t t p://www.mpa-garching.mpg.de/$\tilde{~}$max/minmass.html} (faster from Europe)\\
and from from {\it h t t p ://www.sns.ias.edu/$\tilde{~}$max/minmass.html} (faster from the US).\\
Figures 5 and 6 will print in color if your printer supports it.}
}

% Ju flera kockar, desto saemre soppa.
\vskip 0.5truecm

Max Tegmark$^{1,2}$, 
Joseph Silk$^3$, 
Martin J. Rees$^4$,
Alain Blanchard$^5$, 
Tom Abel$^2$ \&
Francesco Palla$^6$ 

\smallskip
\addr{$^1$Max-Planck-Institut f\"ur Physik, 
F\"ohringer Ring 6,D-80805 M\"unchen; max@mppmu.mpg.de}\\
\addr{$^2$Max-Planck-Institut f\"ur Astrophysik, 
%: Karl-Schwarzschild-Str. 1, D-85740 Garching}\\
Karl-Schwarzschild-Str. 1, D-85740 Garching; abel@mpa-garching.mpg.de}\\
\addr{$^3$Departments of Astronomy and Physics, and 
Center for Particle Astrophysics,}\\
\addr{University of California, Berkeley,
California 94720; silk@pac2.berkeley.edu}\\
\addr{$^4$Institute of Astronomy, University of Cambridge, 
Cambridge CB3 OHA, UK; mjr@ast.cam.ac.uk}\\
\addr{$^5$Observatoire de Strasbourg, 11 rue de l'universit\'e,
67000 Strasbourg, France; blanchard@cdsxb6.u-strasbg.fr}\\
\addr{$^6$Osservatorio Astrofisico di Arcetri,
Largo E. Fermi, 5 -- 50125 Firenze, Italy; palla@arcetri.astro.it}\\

\vskip 0.7truecm

{\bf Abstract}
\end{center}
\smallskip

The minimum mass that a virialized 
gas cloud must have in order to be able to
cool in a Hubble time is computed,
using a detailed treatment of the chemistry
of molecular hydrogen. 
With a simple model for halo profiles, 
we reduce the problem to that of numerically integrating 
a system of chemical equations.
The results agree well with
numerically expensive 3D simulations,
and our approach has the advantage of
rapidly being able to explore large regions of parameter space.
The minimum baryonic mass $M_b$
is found to be strongly redshift dependent,
dropping from $10^6 M_{\odot}$ at $z\sim 15$ to 
$5\times 10^3 M_{\odot}$ at $z\sim 100$ as molecular cooling 
becomes effective. For $z\gg 100$, $M_b$ rises again,
as CMB photons inhibit $H_2$-formation through the 
$H^-$ channel. Finally, for $z\gg 200$, the $H_2^+$-channel
for $H_2$-formation becomes effective, driving 
$M_b$ down towards $M_b\sim 10^3\Ms$.
With a standard CDM power spectrum with $\sigma_8=0.7$, 
this implies that a
fraction $10^{-3}$ of all baryons may have formed luminous objects by $z=30$,
which could be sufficient to reheat the universe.

\end{titlepage}

%%%%%%%%%%%%%%%%%%%%%%%%%%%%%%%%%%%%%%%%%%%%%%

\section{INTRODUCTION}

\subsection{When did the universe reheat? Observational constraints}

It is now widely accepted that the universe underwent a reheating phase at
some point after the standard recombination epoch at redshift 
$z\approx 10^3$. However, the question of when this happened remains open.
The absence of a Gunn-Peterson trough in the spectra of
high-redshift quasars has provided strong
evidence for the reheating occurring at a redshift $z>5$, since it indicates
that the intergalactic medium (IGM) was highly ionized at lower
redshifts (Gunn and Peterson 1965, Steidel and Sargent 1987, Webb
{\etal} 1992).  
The smallest baryonic objects to go non-linear in a standard cold
dark matter (CDM) model are expected to reionize the IGM at a redshift somewhere
in the range $10 < z < 100$ 
(Bond \& Szalay 1983; Couchman 1985; Couchman \& Rees 1986; 
Fukugita \& Kawasaki 1991; Tegmark, Silk \& Blanchard 1994; 
Tegmark \& Silk 1995; Liddle \& Lyth 1995).
In recent models with baryonic dark matter, reheating and
reionization is predicted to occur at an even higher redshift, typically in the
range $100 < z < 1000$ (Peebles 1987, Gnedin \& Ostriker 1992, Cen {\etal} 1993).

A reheating epoch would have at least two interesting 
classes of effects that may
be measurable today: effects on subsequent structure formation and effects on
the cosmic microwave background radiation (CMB).
Subsequent structure formation would be affected in at least two ways:
\begin{enumerate}
\item
The heating of the IGM up to a higher adiabat would raise the
Jeans mass, thus suppressing the formation of small objects. For instance,
an IGM temperature of $10^5\K$ at a redshift of a few would
suppress the formation of galaxies of mass below $10^{10}\Ms$, thus 
alleviating the ubiquitous problem of theories overpredicting the abundance of
faint galaxies ({\eg} Blanchard {\etal} 1992, 
Kauffman {\etal} 1993; Cole {\etal} 1994). 

\item
If the objects that reheat the IGM
also enrich it with heavy elements, the ability of gas to cool would be
greatly enhanced in the temperature range $10^4\K<T<10^7\K$,
presumably facilitating future structure formation.
%: If the objects that reheat the IGM
%: also enrich it with heavy elements, the ability of gas to cool would be
%: greatly enhanced in the temperature range $10 \K<T<10^7\K$.
%: The enhanced line-cooling by the metals and the subsequent molecule 
%: and dust formation will 
%: presumably facilitate future structure formation.

\end{enumerate}
The CMB would be affected in at least three ways:

\begin{enumerate}

\item
Hot ionized IGM would cause spectral distortions which might violate the
stringent limits on the the Compton $y$-parameter (Mather {\etal} 1994).
This is a problem mainly for BDM models (Tegmark \& Silk 1995).

\item
Spatial fluctuations on angular scales below a few degrees may be suppressed,
while fluctuations on larger scales would remain fairly unaffected. 
Therefore a comparison of the results of current and future 
degree scale experiments with those of COBE (Smoot {\etal} 1992) 
constrains the reionization epoch. 

\item
New spatial fluctuations 
will be generated on smaller angular scales, through the so called 
Vishniac effect (Vishniac 1988, Hu {\etal} 1994). 
The current upper limit on CMB fluctuations on the 1 arcminute scale
(Subrahmanyan {\etal} 1993) places constraints 
on some reheating scenarios. 

\end{enumerate}
In other words, with the recent surge in CMB experiments and the 
considerable numerical,
theoretical and observational results on structure formation, 
the thermal history of the universe is now coming
within reach of our experimental probes. 
In view of this, it is
very timely to theoretically investigate the nature of the reheating epoch in
greater detail, and investigate the properties of the objects that caused it. 
In this paper we will focus on two of the most basic attributes
of these first objects: their mass and their formation redshift.  
Hence the goal
is to derive the mass-redshift distribution of the very 
first objects that might be able to reheat the universe, and
thus set the stage for all subsequent cosmological events. 

\subsection{What does theory predict?}

In both CDM and BDM models of structure formation, the first objects 
predicted to go nonlinear are the smallest ones. 
The crucial question is if cooling will allow the baryonic clouds to
dissipate their kinetic energy and 
collapse more than the dark matter, to eventually become self-gravitating and
form an interesting object like a galaxy, a V.M.O. or a black hole
(see {\eg} Binney 1977, Rees and
Ostriker 1977; Silk 1977; White and Rees 1978; 
Araujo \& Opher 1988, 1989, 1991). 
For low mass objects, the smaller they are, the less efficiently they 
dissipate energy and cool.
Thus a detailed treatment of gas-dynamical processes will predict a 
characteristic mass scale $M_c$ such that objects with $M > M_c$ 
can cool rapidly, whereas smaller lumps will merely remain 
pressure supported and not form anything luminous.
In other words, $M_c$ is the mass scale of the first luminous objects.

Fortunately, making a theoretical estimate of $M_c$ is much 
simpler than the corresponding problem for present-day structure formation.
Today there are large uncertainties in both the metal abundance of the
intergalactic medium (IGM), which affects cooling rates, 
and in the UV background, which affects ionization rates and 
molecular chemistry. Before the first structures formed, there 
was by definition neither metals nor UV background. 

% Tom: Changed
The problem has recently been treated 
realistically 
% at the Laboratoryfor Computational Astrophysics 
using a multi-fluid 3D cosmological
hydrodynamics code which evolves not only the dark,  and baryonic
matter, but also tracks the non-equilibrium chemistry of 9 species,
including hydrogen molecules (Abel 1995; Anninos {\etal} 1996;
Norman {\etal} 1996).  
The main obstacle to this program is computational expense,
because of the large dynamical range involved.
As a complement to such heavy computations, it is thus worthwhile 
to attack the problem with various approximate 
techniques that are fast enough to run many times, thereby exploring 
all of parameter space and finding out which parameter choices and 
initial conditions merit more detailed numerical studies. 
This is the purpose of the present paper.
One such approximate method is that of 
Haiman, Thoul \& Loeb  (1996, hereafter  HTL96),  which numerically
follows the  growth if an isolated density peak that is spherically
symmetric. Although the first structures  to  collapse in CDM  are
typically  sheet-like rather than spherically symmetric, this model
nonetheless illustrates which physical  processes are likely to be the
most  important  in the  full 3D  case.  Since the approach of HTL96
involves   numerically  integrating  a  partial  differential equation
(separately tracking a large number of  spherical shells), it is still
fairly time-consuming, and results are presented for only 24 points in
the $M-z$ plane (see \fig{zmFig}). In  this  paper, we use  a still
simpler approach, involving  nothing but ordinary  differential
equations, which turns out to reproduce the results of HTL96 quite well.
% EndTom
The resulting code is so fast that we can run it thousands of times, 
thereby finding the curve in \fig{zmFig} 
that delimit collapsing 
objects from non-collapsing ones, and study how this curve depends on
cosmological parameters such as $\Omega$, $\Omega_b$ and $h$. 

For the various CDM-based scenarios, the first interesting objects will
turn out to  be rare peaks in the
Gaussian random field of mass between $10^4$ and $10^7\Ms$, at redshifts in the
range $20\simlt z\simlt 100$. 
At these redshifts, the initial IGM temperature is considerably lower than the
virial temperatures in question, so the baryons will initially 
collapse together with
the dark matter. 
These first objects
will have virial temperatures between a few 
hundred and a few thousand degrees, 
which means that the main coolant will be molecular hydrogen.
(Line cooling by hydrogen and helium is negligible for $T\ll 10^4\K$, and 
lithium hydride
and other less abundant molecules become dominant only when $T\ll 500\K$.)
In Section 2, accurate expressions for $H_2$ cooling are presented, and  
it is shown that the pre-collapse $H_2$ abundance 
is typically too low for
the clouds to cool significantly in a Hubble time.
The fate of a virialized lump thus depends crucially on its ability to 
rapidly produce more $H_2$, which is the topic of Section 3. 
Our simple model for the evolution of density 
and temperature is presented in Section 4, and the numerical results are
described in Section 5. The results and their cosmological implications are
discussed in Section 6.

%%%%%%%%%%%%%%%%%%%%%%%%%%%%%%%%%%%%%%%%%%%%%%%%%%%%%%%%%%%%%%%%%%%

\section{COOLING BY MOLECULAR HYDROGEN}

How much molecular hydrogen is needed for a gas cloud to be able to cool in
a Hubble time? This question will be answered in the present section.
The atomic physics of molecular hydrogen cooling has been studied extensively by
many authors, {\eg} Lepp \& Shull (1983). 
An excellent review of what will be needed here is given by
Hollenbach and McKee 1979 (hereafter HM), who also provide a number of useful
analytical fits to various numerical results. 

When an $H_2$ molecule gets rotationally or vibrationally excited through a
collision with an $H$ atom or another $H_2$ molecule, there are two competing
channels through which the ensuing de-excitation can occur. Either the
de-excitation is radiative, which amounts to cooling, or it is collisional, in
which case there is no net energy loss from the gas. When the density $n$ is very
low, the former channel dominates. In this case, the hydrogen molecules spend most of
their time in the ground state or in the $J=1$ rotational state (whose radiative
decay to the ground state is forbidden since $H_2$ has no dipole moment), and
collisional excitations are for all practical purposes instantly followed by a
radiative decay. Thus in the low density limit, the energy loss per unit volume is
proportional to $n^2$. When the density $n$ is very high, on the other
hand, collisions dominate. Thus to a good approximation, the distribution of
molecules in various states is the the Boltzmann distribution of LTE, local
thermal equilibrium, and the energy loss per unit volume is only linear in $n$.
The border between ``high'' and ``low" density is roughly the function $\nc$
defined below. It is temperature-dependent, but lies between $10^3$ and
$10^4\cm^{-3}$ for our regime of interest, $10^2\K<T<10^3\K$. 
A just virialized gas cloud has an overdensity of about
$18\pi^2\approx 178$, {\ie},
a hydrogen density 
\beq{DensityEq}
n\approx 23\>\cm^{-3} \left({h^2\Omega_b\over 0.015}\right) z_{100}^3,
\eeq
so during the early stages of collapse, we are well into the low-density regime
for our parameter range of interest. (Here and throughout this paper we assume a
Helium abundance of $24\%$ by mass.)

Since the fraction $f$ of hydrogen in molecular form will be quite low in our
application (typically below $10^{-3}$), we can neglect $H_2 - H_2$ collisions
and the formulas of HM reduce to the following:
The cooling rate is
\beq{CoolingEq1}
L \approx {\Lrlte\over 1 + \nc/n},
\eeq
where the critical density is
\beq{ncEq}
\nc \equiv {\Lrlte\over\Lrlow}n,
\eeq
which depends only on temperature, not on $n$.
Here the cooling rate in LTE is 
\beq{LrlteEq}
\Lrlte \approx {1\over n} 
\left\{
\left({9.5\tento{-22}T_3^{3.76}\over 1+0.12
T_3^{2.1}}\right)e^{-(0.13/T_3)^3} 
+ 3\tento{-24}e^{-0.51/T_3}
\right\}
\ergs\,\cm^3\,\s^{-1},
\eeq
whereas the cooling rate in the low-density limit is 
\beq{LrlowEq}
\Lrlow \approx
{5\over 4}\gamma_2 (E_2-E_0) e^{-(E_2-E_0)/kT}
+ {7\over 4}\gamma_3 (E_3-E_1) e^{-(E_3-E_1)/kT}.
\eeq
Here $T_3=T/1000\K$, $E_J =  J(J+1) E_1/2$, 
where  $E_1/k \approx 171\K.$ Thus
$(E_2-E_0)/k = (3/5)(E_3-E_1)/k =  3 E_1/k  \approx 512\K.$ 
$\gamma_2$  and
$\gamma_3$ are the collisional  de-excitation rates from the $J=2$ and
$J=3$  rotational  levels.   The  rates   for collisional   quadrupole
de-excitation $J\to J-2$ due to impact of hydrogen  atoms are well fit
by (HM)
\beq{gammaEq}
\gamma_J(T) = 
\left({10^{-11} T_3^{1/2}\over 1 + 60 T_3^{-4}} + 10^{-12} T_3\right)
\left(0.33 + 0.9\exp\left[-\left({J-3.5\over 0.9}\right)^2\right]\right)
\cm^3\,\s^{-1}.
\eeq
\Eq{LrlowEq} assumes an ortho-$\H_2$ to para-$H_2$ ratio of 3:1.
The first term gives the cooling contribution of para-$\H_2$ and the
second that of ortho-$\H_2$. Abel {\etal} (1996b) show that for $\H_2$
formation by the gas phase reactions discussed in the following
section, the interconversion mechanism,
\beq{OrthoParaEq}
\H_2(ortho) + \H^+ \to \H_2(para) + \H^+
\eeq
will be fast enough to convert all ortho-H$_2$ to para-H$_2$.  Hence
the appropriate cooling rate is given by the first term in
\eq{LrlowEq} multiplied by four, {\ie}, \eq{LrlowEq} is replaced
by simply
\beq{LrlowEq2}
\Lrlow \approx
5\gamma_2 (E_2-E_0) e^{-(E_2-E_0)/kT}.
\eeq
Defining the {\it cooling timescale} $\tc\equiv
T/\dot T$, we thus obtain
\beq{MolCoolEq}
\tc \approx 48200\years\left(1+{10 T_3^{7/2}\over 60 + T_3^4}\right)^{-1}
e^{512\K/T}(f n_1)^{-1},
\eeq
where $n_1\equiv n/1\,\cm^{-3}$.
Let us define the {\it Hubble timescale}  $\th$ at a redshift $z$ as the age of
the universe at that redshift. Then for $\Omega=1$, 
\beq{HubbleTimeEq}
\th\aet{6.5}{6}\years\>h^{-1}z_{100}^{-3/2}.
\eeq
Since the primordial gas clouds in which we are interested have just virialized,
the Hubble timescale is of the same order as 
% in fact, they differ by a factor sqrt(18\pi^2)
the {\it gravitational
timescale} 
$\tg \equiv (\rho G)^{-1/2}$, the timescale on
which collapse would proceed if the temperature where lowered and the clouds 
lost their pressure support. 
Thus the future of a newly formed gas cloud is crucially dependent on the
ratio ${\tc/\th}$ (Rees \& Ostriker 1977).
If $\tc\ll\th$, the gas cloud will rapidly cool and begin a nearly
free-fall collapse, whereas if $\tc\gg\th$, the cloud will remain pressure
supported and fairly stationary until much lower redshifts.
$\tc = \th$ for 
\beq{fCritEq}
f \approx 0.00016
\left({h\Omega_b\over 0.03}\right)^{-1}z_{100}^{-3/2}
\left(1+{10 T_3^{7/2}\over 60 + T_3^4}\right)^{-1} e^{512\K/T}.
\eeq
This critical $H_2$ fraction is plotted in \fig{tfFig}, 
as a function of
temperature. It is seen that the $H_2$ fraction required exceeds typical
initial abundances ($\sim 10^{-6}$) for all redshifts $z<200$ when $T<10^4\K$. Thus 
our low-mass, high-redshift clouds can cool and collapse only if additional
$H_2$ is produced (unless $T\simgt 10^4\K$, in which case hydrogen line cooling 
will be effective). 

In the following section, we will compute how much additional $H_2$ will be
produced, and discuss the conditions for when sufficient cooling will
indeed occur. 

%%%%%%%%%%%%%%%%%%%%%%%%%%%%%%%%%%%%%%%%%%%%%%
\begin{table}
\begin{tabular}{|l|l|l|} 
\hline
Reaction			&Rate $k$ $[\cm^3/\s]$		&Reference\\
\hline
%$\H^+ +e^-\mapsto\H +h\nu$	&$\recrate\approx 1.88\tento{-10} T^{-0.64}$	&Seaton (1959)\\
$\H^+ +e^-\mapsto\H +h\nu$	&$\recrate\approx 1.88\tento{-10} T^{-0.64}$	&Hutchins (1976)\\
$\H+e^-\mapsto\H^-+h\nu$	&$\molrate\approx 1.83\tento{-18} T^{0.88}$	&Hutchins (1976)\\
$\H^-+\H\mapsto\H_2 + e^-$	&$\goodrate\approx 1.3\tento{-9}$		&Hirasawa (1969)\\
%$\H^-+\H\mapsto\H_2 + e^-$	&$\goodrate\approx 1.35\tento{-9}$		&Hutchins (1976)\\
$\H^++\H\mapsto\H_2^+ +h\nu $	&$\molratetwo\approx 1.85\tento{-23}T^{1.8}$	&Shapiro \& Kang (1987)\\
$\H_2^++\H\mapsto\H_2+\H^+$	&$\goodratetwo\approx 6.4\tento{-10}$		&Karpas {\etal} (1979)\\
% \hline
% $\Li+\H\mapsto\Li\H+h\nu$	&$10^{-17}$			&Dalgarno \& Lepp (1987)\\
\hline
\hline
Reaction			&Rate $k$ $[1/\s]$		&Reference\\
\hline
$\H^-+h\nu\mapsto\H + e^-$	&$\badrate\approx 0.114\Tg^{2.13}e^{-8650/\Tg}$&Appendix A\\
$\H_2^+ +h\nu\mapsto\H + H^+$	&$\badratetwo\aet{6.36}{5} e^{-71600/\Tg}$&Appendix A\\   % 30750
$e^- +h\nu\mapsto e^- + h\nu$	&$\comprate\approx 4.91\tento{-22}\Tg^4$&\\
%$\H^-+h\nu\mapsto\H + e^-$	&$\badrate\approx 0.278\Tg^{2}e^{-8750/\Tg}$&Hirasawa (1969)\\
%$\Li\H+h\nu\mapsto\Li+\H$	&				&Kirby \& Dalgarno (1978)\\
%$\H_2+h\nu\mapsto\H+\H$	&$0.01B_{\lambda_0}$		&Lenzuni {\etal} (1991)\\
\hline
\end{tabular}%
%\caption{Rates for photodissociation by CMB}
\caption{Reaction rates used (all temperatures in Kelvin)}
\label{RateTable}
\end{table}

%%%%%%%%%%%%%%%%%%%%%%%%%%%%%%%%%%%%%%%%%%%%%%
%%%%%%%%%%%%%%%%%%%%%%%%%%%%%%%%%%%%%%%%%%%%%%
%%%%%%%%%%%%%%%%%%%%%%%%%%%%%%%%%%%%%%%%%%%%%%%%%%%%%%%%%%%%%%%%%%%
\section{PRODUCTION OF MOLECULAR HYDROGEN}

\begin{figure}[phbt]
\centerline{\rotate[r]{\vbox{\epsfysize=16cm\epsfbox{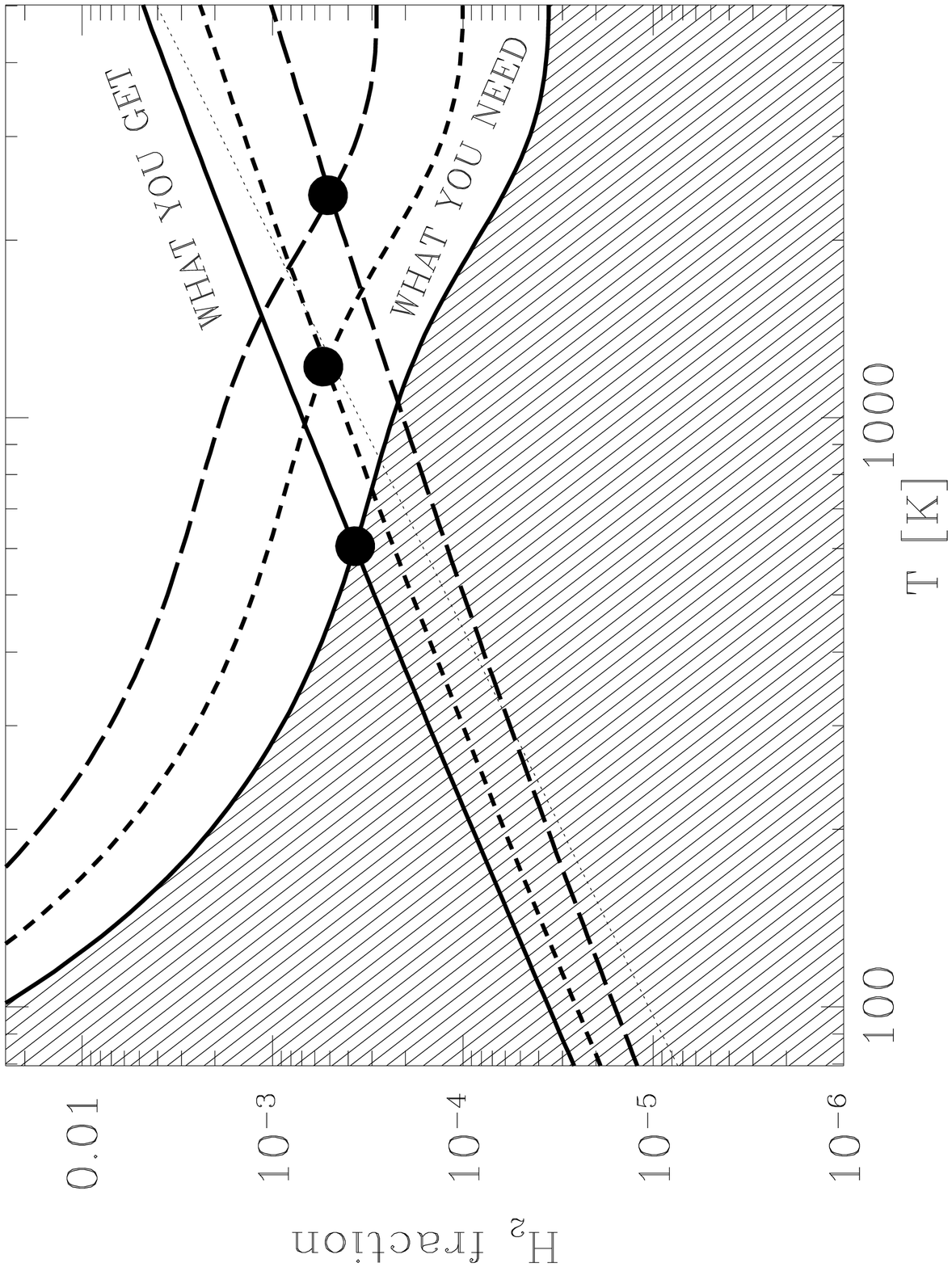}}}}
%\centerline{{\vbox{\epsfxsize=16cm\epsfbox{tf.ps}}}}
\caption{Molecular fraction needed and molecular fraction produced.}
\label{tfFig}
The solid, short-dashed and long-dashed lines correspond to
lumps virializing at $\zvir=100$, 50 and 25, respectively. 
Only clouds above the downward sloping lines 
(outside the shaded region for $\zvir=100$) can cool in a Hubble time.
The upward-sloping lines show the molecular fraction produced in a 
local Hubble time, so the minimum temperature needed for collapse is that
where the pair of curves cross (solid dots --- lower $\zvir$ require
higher virial temperature).
Electron depletion is the limiting factor above the thin dotted
line, so we see that for $z\simgt 50$, the results are rather 
independent of the initial ionization fraction.
\end{figure}

How much molecular hydrogen will be produced in a Hubble time?
In hydrogen of density $n = n[H] + n[H^+] + 2n[H_2]$ at temperature $T\simlt 10^3\K$,
the ionization fraction $\x\equiv n[H^+]/n$ 
and the molecular fraction $\f\equiv n[H_2]/n$ evolve as
%: CHANGED TO:
%:and the molecular fraction $\f\equiv n[H_2]/n$ in primordial gas of slowly 
%:changing density ($dn[H^+]/dt = d(nx)/dt \approx ndx/dt$) evolve as
\beq{xdotEq}
\dot\x = -\Lrec n\x^2,
\eeq
\beq{fdotEq}
\dot\f = \Lmoltot n (1-x-2f) \x.
\eeq
Collisional ionization of $H$ atoms as well as collisional dissociation of $H_2$
is completely negligible at such low temperatures. 
% The recombination rate is well approximated by (Seaton 1959)
% $$\Lrec\aet{2.26}{-12}T_3^{-0.64}\cm^3 s^{-1}.$$
At the low densities in question, $H_2$ is formed mainly via the reaction
$H+e^- \to H^- + h\nu$ at the rate $\Lmol$,
after which one of the following two things happen to the $H^-$
almost instantaneously: 
\begin{itemize}
\item Molecular hydrogen is produced through the
reaction $H + H^- \to H_2 + e^-$, at the rate $\goodrate$.
\item The $H^-$ gets destroyed by a CMB photon, at the rate $\badrate$. 
\end{itemize}
Thus the effective rate of $H_2$-formation is 
$\Lmol\goodrate/[\goodrate +\badrate/(1-x)n]$. 
Since $\Tg\propto 1+z$, the exponential term in $\badrate$ 
effectively makes $H_2$-production through the $H^-$-channel 
impossible for $z\gg 200$. 
A second, less effective channel for molecule formation 
is the slow reaction $H^+ +H\to H_2^+ + h\nu$ at the rate
$\molratetwo$, followed almost immediately by either 
$H_2^+ + H \to H_2 + H^+$
at the rate $\goodratetwo$ or photodissociation at the rate
$\badratetwo$, thus producing $H_2$ at the 
net rate $\molratetwo\goodratetwo/[\goodratetwo+\badratetwo/(1-x)n]$.
This channel works up to higher redshifts, but since 
$\molratetwo\ll\molrate$, it becomes important 
only for lumps with virialization redshifts $\zvir\gg 100$.
In our calculations further on, we use the exact rate, {\ie},
\beq{MolRateEq}
\Lmoltot = \left[{\goodrate\over\goodrate+\badrate/(1-x)n}\right]\molrate
         + \left[{\goodratetwo\over\goodratetwo+\badratetwo/(1-x)n}\right]\molratetwo.
\eeq 
Although we integrate the above-mentioned 
chemical equations numerically
in our analysis, a number of the features of the solutions
can be readily understood
from the following elementary observations.
First note that \eq{xdotEq} is independent of $f$, since 
the electrons act only as a catalyst in the reactions that
produce $H_2$. Since the right-hand-side of \eq{xdotEq} 
is not linear but
quadratic in $x$, the residual ionization fraction 
decays much slower than
exponentially. In the absence of cooling, $T$ and $n$ 
will remain roughly constant
in the pressure-supported cloud, and the solution will be 
\beq{xSolEq}
x(t) = {x_0\over 1+x_0 n\Lrec t}\>,
\eeq 
{\ie}, $\x\to 0$ only as $1/t$, where $\Lrec$ is the 
recombination rate.\footnote{
To obtain better accuracy when $z\sim 10^3$, we use the more complicated
rate equations given in Peebles (1993, \S 6) in place of the rate 
$k_1$ from Table 1 in our numerical runs. 
}
Substituting this into 
\eq{fdotEq}, we see that $f\to 1$ as
$t\to\infty$, {\ie}, all hydrogen would become molecular if we waited long enough.
With parameters in our range of interest, however, $f$ will remain much less than
unity for many Hubble times. Thus taking $1-\x-2\f\approx 1$,
\eq{fdotEq} has the solution
\beq{fSolEq}
f(t) = f_0 + {\Lmoltot\over\Lrec}\ln(1 + x_0 n\Lrec t)
\eeq
when $\Lmoltot$ is roughly constant (it will be roughly constant
except at $z\sim 300$ and $z\sim 100$, which is when 
the two radiative dissociation processes go from being dominant to negligible).
Thus the time evolution separates into two distinct regimes:
$x_0 n \Lrec t\ll 1$ and $x_0 n \Lrec t\gg 1$. 
In the 
first regime, the residual
ionization remains roughly constant, and molecules get produced at a constant
rate. In the second regime, 
electron depletion becomes a serious problem, and the
molecular fraction grows only logarithmically with time. 
Since the factor $1/(x_0 n k)$ is simply the recombination timescale, 
we can rephrase this result as stating that
the molecule fraction produced is 
$f-f_0 = (\Lmoltot/\Lrec)\ln(1 + N_{rec})$, where $N_{rec}$ 
is the number of recombination times elapsed.
The transition occurs after about one recombination time $(N_{rec}\approx 1)$,
{\ie}, when 
\beq{DepletionEq}
f \approx f_c \equiv {\Lmoltot\over\Lrec} 
\aet{3.5}{-4}T_3^{1.52}
\left[1+7.4\tento{8}n_1^{-1}(1+z)^{2.13} e^{-3173/(1+z)}\right]^{-1}
\eeq
for $z\ll 300$,
a value that is independent of the initial ionization fraction $\x_0$.
The factor in square brackets corresponds to photodissociation
of $H^-$, and can be ignored for $z\simlt 100$. 
\Fig{tfFig} shows $f(\th)$ as a function of $T$ for 
$x_0 = 3\times 10^{-4}$, together with
$f_c(T)$. As can be seen, we typically have $f(\th) > f_c$ for 
$\zvir\gg 50$,  {\ie}, we are
well into the electron depletion regime, 
which means that the final molecule abundance $f$
is rather insensitive to the initial ionized fraction $\x_0$
and approximately given by \eq{DepletionEq}.

\Fig{tfFig} also shows that the three solid dots 
almost line up horizontally. In other words, the 
molecular fraction in clouds that just barely manage to collapse
(where the molecular hydrogen fraction produced 
within a Hubble time is just enough to make it cool in a Hubble time)
is almost independent of the virialization redshift for
$25 \simlt \zvir \simlt 100$. 
Since the virial temperature of a collapsing
cloud is determined only by its mass and its virialization redshift,
this implies that any cloud with
a molecular hydrogen fraction $\sim5\tento{-4}$ is able to
cool within a Hubble time. 
We can summarize this with the following useful rule of thumb:
If the virial temperature is high enough to produce a molecular 
hydrogen fraction of order $5\tento{-4}$,
then the cloud will collapse. This explains
the rather constant slope in figures~\ref{ztFig}
and~\ref{zmFig} for 
$20\simlt\zvir\simlt 80$.

\section{EVOLUTION OF DENSITY AND TEMPERATURE}

In this section, we describe our simple model for 
how the gas density and temperature evolve in
an overdensity that grows, goes nonlinear and virializes.
Section 4.1 refers mainly to the dark matter --- the late stages of
the density evolution of the baryons is discussed in 4.2 and 4.3.

\subsection{The density}

\begin{figure}[phbt]
%\centerline{\rotate[r]{\vbox{\epsfysize=16cm\epsfbox{zn.ps}}}}
\centerline{{\vbox{\epsfxsize=14cm\epsfbox{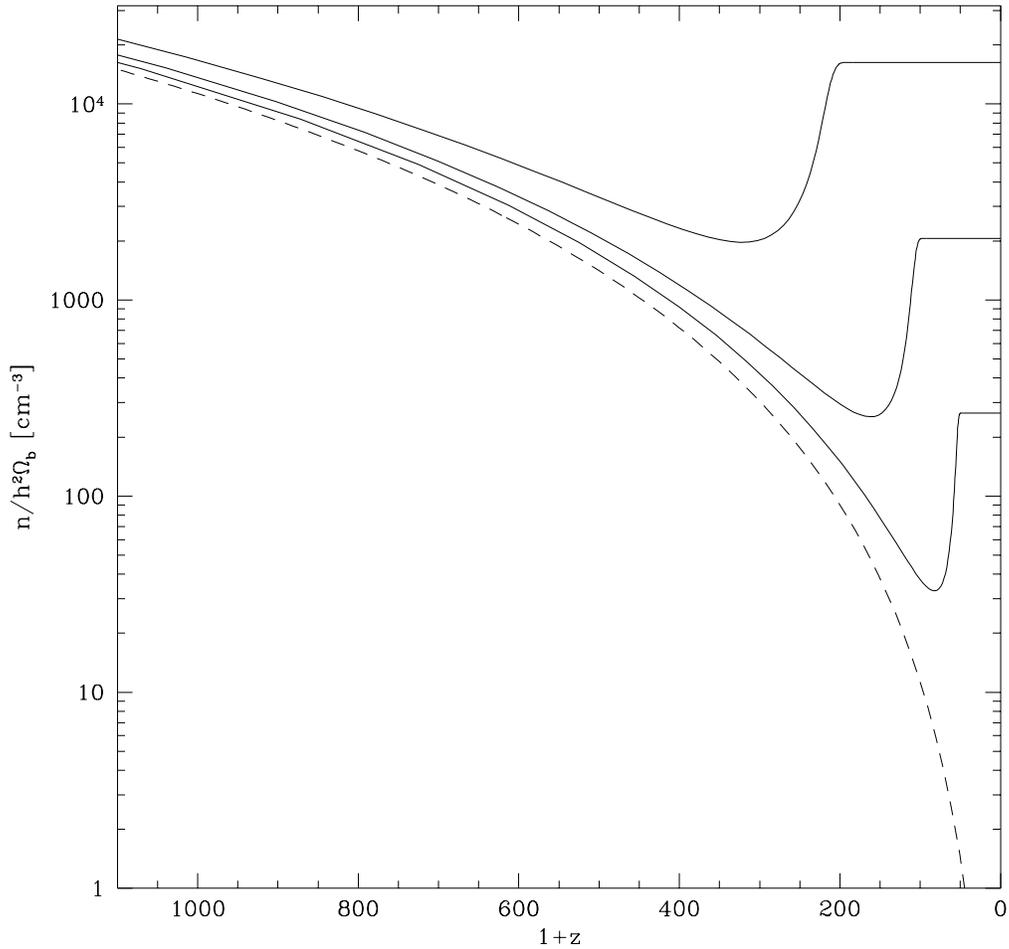}}}}
\caption{Model for density evolution.}
Our model for the evolution of the baryon number density $n(z)$
is shown for models with three different virialization redshifts
$\zvir$, for the case of negligible pressure. 
$n$ first decreases slower than the background 
density (dashed line) according to linear theory, then 
increases again as the lump collapses and virializes, and finally
reaches the virial plateau value of  
$18\pi^2$ times the background density when $z=\zvir$.
\label{znFig}
\end{figure}

Early on, while $z\gg\Oz^{-1}$, space is approximately flat and 
the Friedmann equation has the approximate solution
\beq{EarlyFriedmanEq}
a(t)\propto t^{2/3}
\eeq
regardless of the values of $\Oz$ and the cosmological constant $\Lz$. 
If an $\Omega=1$ universe has a completely uniform density
$\rho$ except for a ``top hat" overdensity, 
a spherical region where the density
is some constant $\rho'>\rho$, then this top hat region will gradually begin to
expand slower than the rest of the universe, stop expanding and recollapse to a
point. By Birkhoff's theorem, the radius of this region will evolve according to
the Friedmann equation, but with some $\Omega>1$. 
It is well known that the overdensity 
\beq{deltaDefEq}
\delta\equiv{\rho'\over\rho} - 1
\eeq
evolves as 
\beq{deltaEvolEq}
(1+\delta) = 
{9\over 2}{(\alpha - \sin\alpha)^2\over(1-\cos\alpha)^3}
= 1 + {3\over 20}\alpha^2 + O(\alpha^3),
\eeq
where the parameter $\alpha$, the ``development angle" 
is related to the redshift through
\beq{DevAngleEq}
{1+\zvir\over 1+z} = \left({\alpha - \sin\alpha\over 2\pi}\right)^{2/3}
= {\alpha^2\over (12\pi)^{2/3}} + O(\alpha^{8/3}).
\eeq
Here $\zvir$ is the redshift at which the top hat would collapse to a point. In
reality, an overdense region would of course not collapse to a point
(and form a black hole).
Since it would
not be perfectly spherically symmetric,  
collisionless
dark matter particles would
mostly miss each other as they whizzed past the central region and out again on
the other side, eventually settling down in some (quasi-) equilibrium 
configuration known as the virial state. 
For baryons, gas-dynamical processes become important, and pressure eventually 
halts the collapse at some density $\rho_p$
as discussed in Section~\ref{LoebSec} below.
Strictly speaking, virial states are not
stable over extremely long periods of time, and their density is certainly not
uniform. 
For a virialized lump, often referred to as a ``halo", 
a typical density profile peaks 
around some constant value in its core and falls off like $1/r^2$ over
some range of radii. Nonetheless, halos are often said to have a ``typical"
density 
\beq{rhovirEq}
\rhovir \approx 18\pi^2\rho_0(1+\zvir)^3,
\eeq
which is a useful rule of thumb. 
Thus in the top-hat collapse model, density in the perturbed region is assumed to
evolve as in \fig{znFig}:
the density starts out decreasing almost as fast as the background
density $\rho$, with 
$$\delta \propto (1+z)^{-1}$$ 
early on, just as in linear
theory, but gradually stops decreasing and increases radically as $z$ approaches
$\zvir$. 
It never increases past the 
virial value $\rhovir$ 
or the pressure-determined value
$\rho_p$, whichever is smaller, 
but stays at that density for all $z<\zvir$.
The main motivation for the use of the Lagrangian code in 
HTL96 was to provide a more realistical modeling of the spatial 
structure of the halo. We use the simple top-hat approximation
instead, for the following reasons:
\begin{itemize}
\item It requires much less computer time.
\item It reproduces the results of HTL96 fairly well.
\item The spherical symmetry assumption of HTL96 is probably somewhat
inaccurate anyway, since n-body simulations have demonstrated that 
the first collapsed structures tend to be sheetlike pancakes rather 
than spherically symmetric.
\end{itemize}
In defense of the spherical symmetry assumption, 
very rare peaks in a random field (which might correspond to the
very first objects) are typically almost spherically symmetric 
(Bardeen {\etal} 1986). More importantly, 
since the virial temperatures in our application are typically only  
slightly higher than the pre-collapse gas temperatures,
none of our conclusions should be very sensitive to 
the actual way in which the cloud gets to its virial configuration, 
such as whether it first passes through an intermediate pancake-like
configuration or not.

Unfortunately, $\alpha$ cannot be eliminated from the equations that relate
$\delta$ and $z$ by using elementary functions. 
For this reason, we use the following fit to the density
evolution $\rho(z) = \rho_0[1+\delta(z)]$, which is
accurate to about $5\%$ until $z$ is within $10\%$ of $\zvir$
(Tegmark 1994), at which the
density is assumed to start approaching the limiting value 
$\rhovir$ or $\rho_p$ anyway:
\beq{rhoFitEq}
\rho(z)\approx 
\rho_0(1+z)^3
\exp\left[-{1.9A\over 1-0.75
A^2}\right],
\eeq
where 
\beq{Aeq}
A(z) \equiv {1+\zvir\over 1+z},
\eeq
and $\rho_0$ is mean density of the universe today.
We use this fit in our numerical analysis, but never let the density 
exceed the virial value, as shown in \fig{znFig}.

\subsection{The temperature}

The thermal evolution of the gas is dominated by the following processes:
\begin{itemize}
\item
Hydrogen line cooling (as given by \eq{LineCoolEq})
\item 
Cooling by molecular hydrogen (as given by \eq{MolCoolEq})
\item
Compton cooling (as given by \eq{ComptonEq})
\item 
Adiabatic cooling/heating 
(caused by the expansion/compression of the gas)
\end{itemize} 
Bremsstrahlung and helium line cooling are completely negligible at the
low temperatures in which we are interested.
The first three mechanisms simply couple the gas atoms to the radiation field,
which means that they will cause cooling when the gas is 
hotter than the CMB and heating otherwise. In other words, none of 
these mechanisms can make the gas cooler than the CMB temperature,
which at $z=100$ is a few hundred K.\footnote{
Assuming nucleosynthesis abundances, 
cooling by lithium hydride is negligible compared to $H_2$-cooling
unless $T\ll 100\K$ (Puy {\etal} 1993, Puy \& Signore 1995), so we can safely 
neglect lithium chemistry for our application.
}  
In the Compton case,
this is reflected by the fact that the cooling rate is of the form
\beq{ComptonEq}
\left({dT\over dt}\right)_{\rm comp} = \comprate x(\Tg-T).
\eeq
For line cooling, given by (Dalgarno \& McCray 1972)
\beq{LineCoolEq}
\linecoolrate\aet{7.5}{-19}\>\ergs\>\cm^{3}\s^{-1} e^{-118348\K/T} n^2 x(1-x),
\eeq
the CMB temperature is completely irrelevant, since line cooling
only becomes important when $T\gg 10^3\K$, {\ie}, when 
$T\gg\Tg$. For the molecular case, this is included by 
replacing $\molcoolrate(T)$ by the net cooling rate
$\molcoolrate(T)-\molcoolrate(\Tg)$.

The adiabatic contribution is given by the $p\,dV$ work done as the gas 
expands or contracts. In the simple top-hat model of the previous 
section, the density of the lump remains almost uniform until close to the
virialization redshift $\zvir$, so that the adiabatic cooling term is
simply
\beq{AdiabCoolEq}
\left({dT\over dt}\right)_{\rm adiab}
 = {2\over 3} {\dot n\over n} T,
\eeq
where the baryon number density $n\propto\rho$ is given by 
\eq{rhoFitEq}. (The molecular abundances are so small that 
to a good approximation, we can treat the IGM as a $\gamma=5/3$ 
monoatomic ideal gas.)
As $z\to\zvir$, \eq{rhoFitEq} would imply that $T\to\infty$, as the
lump collapses to a point. Instead, the lump is assumed to settle into an
approximately pressure-supported configuration, where a typical gas 
element will obtain the virial temperature $\Tvir$. 
For an overdense lump of total (baryonic and dark) mass $M$ that
stops expanding, recollapses and virializes at 
redshift $\zvir$, this temperature $\Tvir$, which corresponds 
to the gas particles having similar velocities as the dark matter 
particles, is approximately (Blanchard {\etal} 1992)
\beq{TvirEq}
\Tvir = 485\K\>h^{2/3}
\left({M\over 10^4\Ms}\right)^{2/3}\left({1+\zvir\over 100}\right). 
\eeq

\subsection{The effect of gas pressure}
\label{LoebSec}

How high will the typical gas density be in this pressure-supported state?
At redshifts $\gg 100$, the Compton coupling to the CMB via the 
small fraction ($10^{-5}-10^{-3}$) of the electrons that remain ionized is
still so strong that the IGM temperature will be close to that of the CMB,
\beq{TcmbEq}
\Tg\approx 273\K \left({1+\zvir\over 100}\right).
\eeq
As time progresses, the Compton coupling weakens, 
and the IGM begins
to cool below the temperature $\Tg$, cooling adiabatically as $(1+z)^2$. 
Comparing \eq{TvirEq} and \eq{TcmbEq}, 
we therefore see that as long as $M\gg 10^4\Ms$, 
the baryons in the ambient IGM will have a temperature
considerably below $\Tvir$, and begin to fall into
a virial configuration together with the cold dark matter.
However, the gas density can only rise by the large 
factor $18\pi^2$ without problems with pressure support 
if $T\ll\Tvir$ {\it after} the collapse.
Since $T\propto n^{2/3}$ during the adiabatic compression, 
this means that we must have 
$\Tvir \gg (18\pi^2)^{2/3}T\sim 32\,T$ before the collapse
to be able to ignore pressure, and this turns out to be a 
good approximation for the critical masses only
when $\zvir\ll 100$. Otherwise, the condition 
that $T=\Tvir$ after the collapse gives only 
a collapse factor of order $(\Tvir/T_1)^{3/2}$, 
where $T_1$ denotes the temperature
of the uniform background medium at redshift $z=\zvir$.
In other words, 
\eq{rhovirEq} is replaced by 
\beq{rhovirEq2}
\rho_p \approx \rho_0(1+\zvir)^3 \left({\Tvir\over T_1}\right)^{3/2}.
\eeq
We can obtain a more rigorous estimate of the final density 
as follows (Loeb 1996).
Hydrostatic equilibrium after the collapse implies that gravity
is balanced by pressure gradients, {\ie}, that
the gravitational potential $\phi$ and the pressure $p$ are related by
\beq{GradientEq}
\nabla\phi = - {1\over\rho}\nabla p.
\eeq
Integrating this equation along some curve from very far outside the lump
(where $\phi=0$ by definition) to a typical point inside the lump, 
we thus obtain
\beq{PhiEq}
\phi = \int\nabla\phi\cdot d\vr = \int{\nabla p\over\rho}\cdot d\vr.
\eeq
Since the gas has been compressed adiabatically during the collapse to
this state, its pressure and density are related by 
\beq{AdiabEq}
\left({p\over p_1}\right) = \left({\rho\over\rho_1}\right)^{5/3}, 
\eeq
where $p_1$ and $\rho_1$
denote the pressure and density
of the uniform background medium at redshift $z=\zvir$.
Substituting this into \eq{PhiEq}, we obtain
\beq{PhiEq2}
\phi 
%= {1\over\rho_1}\int\left({p\over p_1}\right)^{-3/5}\nabla p \cdot d\vr
= {5\over 2}{p_1\over\rho_1}\int\nabla \left({p\over p_1}\right)^{2/5}\cdot d\vr
= -{5\over 2}{p_1\over\rho_1}\left[\left({p\over p_1}\right)^{2/5} - 1\right].
\eeq
By the ideal gas law, $p_1/\rho_1 = kT_1/m_p$, where $m_p$ is the molecular
weight. Eliminating $(p/p_1)$ using \eq{AdiabEq} and defining 
$\Tvir$ by 
\beq{TvirDefEq2}
{3\over 2}k\Tvir = -{1\over 2}m_p\phi,
\eeq
we thus find the final overdensity inside the lump to be 
\beq{LoebEq}
(1+\delta) = {\rho\over\rho_1} = 
\left[1+{6\over 5} {\Tvir\over T_1}\right]^{3/2},
\eeq
in good agreement with $(1+\delta) = (\Tvir/T_1)^{3/2}$
from \eq{rhovirEq2} considering that the 
factor $1/2$ in the definition of $\Tvir$ in \eq{TvirDefEq2}
was somewhat arbitrary.
In reality, the gas evolution might not be completely adiabatic
during the collapse, because of the above-mentioned
cooling processes. 
\footnote{
Our derivation also neglected entropy
generation due to the thermalization of bulk kinetic energy. 
When an object virializes, the infall kinetic energy
of the gas is thermalized in a virialization shock. Thus some entropy is
generated and the pressure of the gas is higher 
(typically by a factor 1-2)
than predicted by the adiabatic compression. In any
event, this entropy generation would only decrease the above 6/5 by a
factor of order unity and would not change the results substantially.
}
We therefore adopt the following 
procedure in our simulations: the gas density is evolved according to
the top-hat solution until $T$ reaches $\Tvir$. At this point, 
gas pressure is assumed to halt the collapse, and 
the gas density is held constant for the rest of the run.
If the gas overdensity reaches the virial value $18\pi^2$ before 
$T$ reaches $\Tvir$, then the density is held constant 
at this value, and the temperature is raised to $\Tvir$ 
(by assumed shocks).

What happens now, after $\zvir$? 
If the gas is going to be able to collapse further and 
eventually form something like population III stars, 
the baryons must now be able to dissipate energy
rapidly through cooling.
If this is the case, the gas cloud may get dense enough
to become self-gravitating, 
which adds further instability to the system
and may eventually lead to the formation of 
an extremely nonlinear object like a galaxy.
The key question is thus how fast the gas in the lump can cool 
after $\zvir$. This is the topic of the next section.

\clearpage
\section{NUMERICAL RESULTS}

After a lump virializes, one of two things will happen to it:
\begin{itemize}
\item
Enough $H_2$ is produced that it will enter a phase of 
runaway cooling and collapse.
\item
Cooling will be so slow that it will remain 
pressure supported for a Hubble time.
\end{itemize}
In the former case, we will say that the lump {\it collapses}, 
in the latter case that it {\it fails} to collapse. If it fails, 
it will not produce any luminous objects that can reheat the 
IGM, but merely remain as an object resembling a small 
Lyman alpha cloud.
Whether a lump succeeds or fails to collapse of course depends on
cosmological parameters such as $h$, $\Omega$ and $\Omega_b$.
First and foremost, it depends strongly on the parameters
$M$ and $\zvir$. In this section, we first give an operational
definition of what we mean by collapse, and then evolve a large number of 
lumps numerically to see for which parts of parameter space they 
manage to collapse, summarized in 
figures~\ref{ztFig} and~\ref{zmFig}.

\begin{figure}[phbt]
\centerline{\rotate[r]{\vbox{\epsfysize=16cm\epsfbox{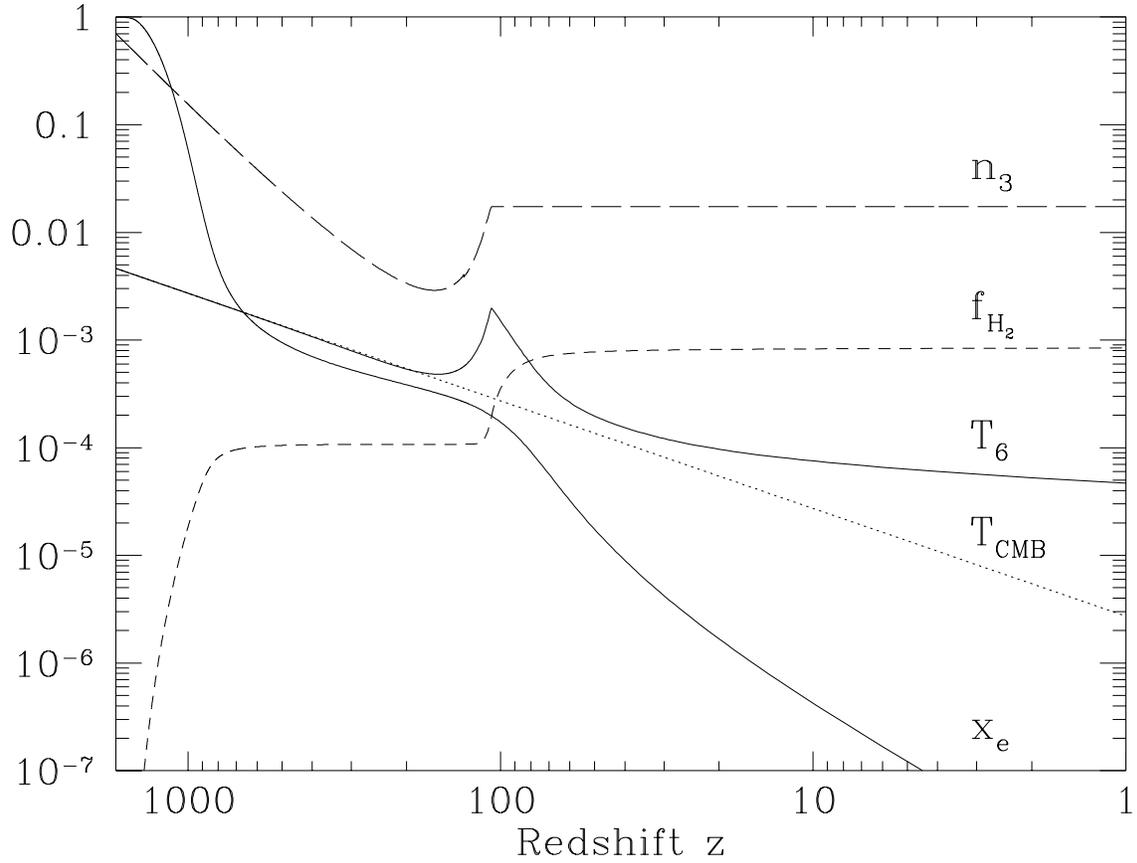}}}}
\caption{
Lump evolution.
}
The time-evolution of gas in a lump is shown for
$\zvir=100$, $\Tvir=2000\K$,
$h=0.5$, $\Omega=1$ and $\Omega_b=0.06$. From 
top to bottom on the right side, the curves show
the number density $n$ in units of $10^3\cm^{-3}$, the 
molecular fraction $f$, the temperature $T$
in units of $10^6K$, the CMB temperature in the same 
units and the ionization 
fraction $x$.
\label{zfFig1}
\end{figure}

\begin{figure}[phbt]
\centerline{\rotate[r]{\vbox{\epsfysize=16cm\epsfbox{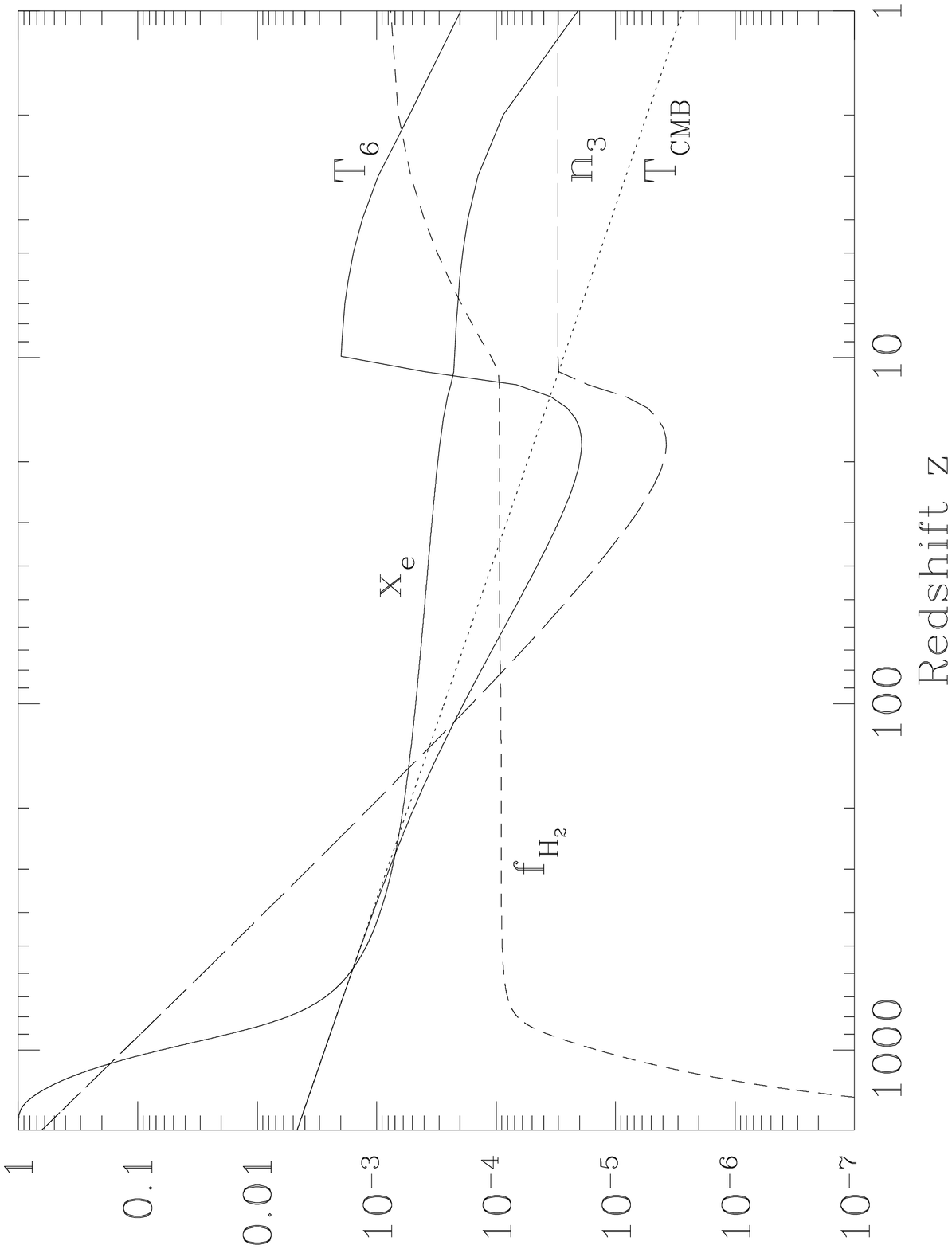}}}}
\caption{
Lump evolution.
}
Same as previous figure, except that $\zvir=10$.
\label{zfFig2}
\end{figure}

\begin{figure}[phbt]
\vskip-4.0cm
\centerline{{\vbox{\epsfxsize=16cm\epsfysize=12cm\epsfbox{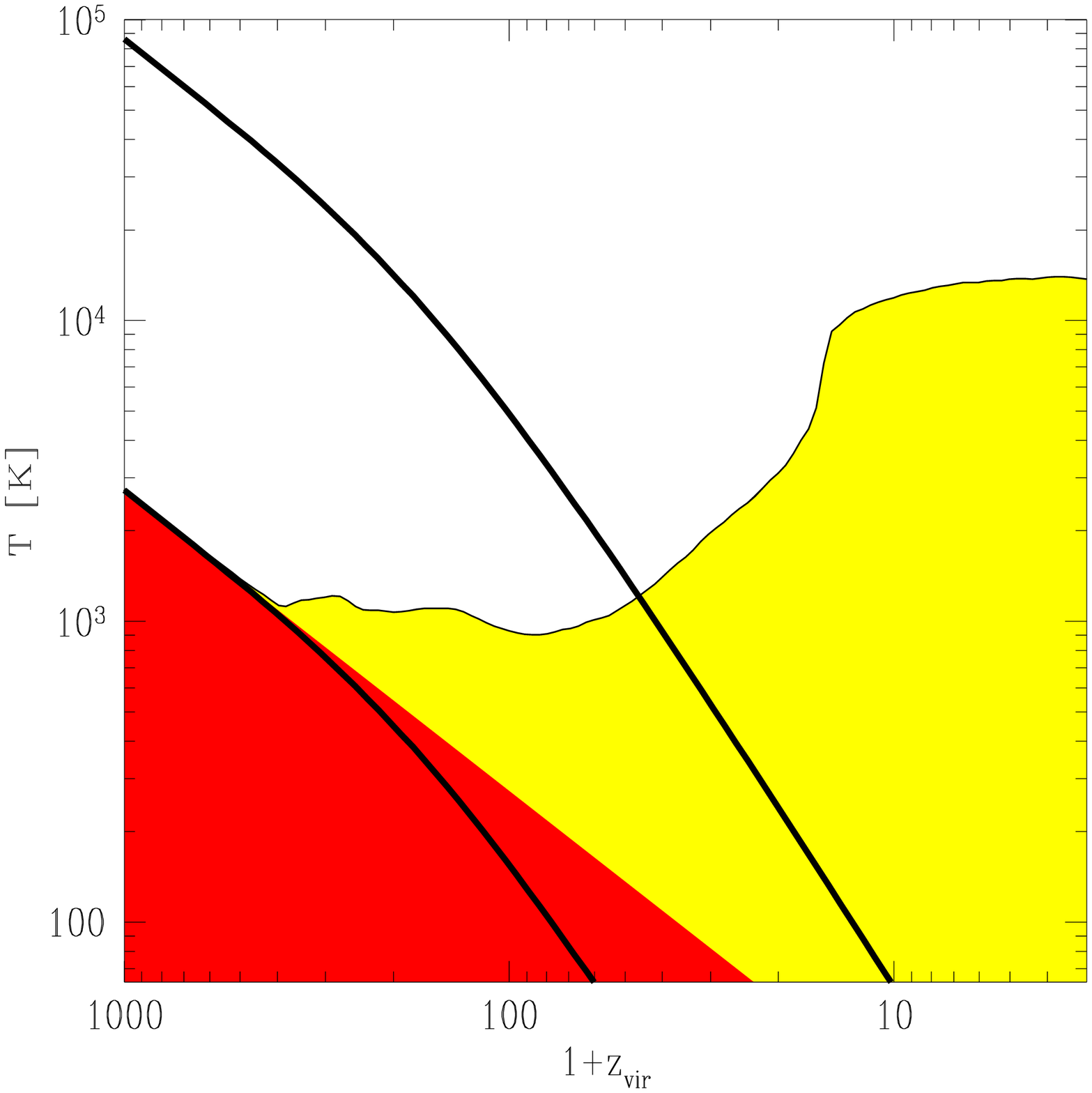}}}}
%\centerline{{\vbox{\epsfxsize=15cm\epsfysize=15cm\epsfbox{zt.ps}}}}
%\vskip-2cm
\caption{The minimum virial temperature needed to collapse.}
The minimum $\Tvir$ for which collapse succeeds is 
plotted at a function of virialization redshift 
for standard CDM 
($\Omega=1$, $\Omega_b=0.06$, $h=0.5$).
%and BDM ($\Omega=\Omega_b=0.15$, $h=0.8$ -- light solid line). 
Only lumps 
whose parameters $(\zvir,\Tvir)$ lie above the shaded area can collapse and
form luminous objects.  
The dark-shaded region is that in which no radiative 
cooling mechanism whatsoever
could help collapse,
since $\Tvir$ would be lower than the CMB temperature.
The solid curves show the temperature evolution of the
uniform IGM and $(18\pi^2)^{2/3}$ times this value,
so above the upper line, gas can 
attain the virial overdensity without problems 
with pressure support.
\label{ztFig}
\end{figure}

\begin{figure}[phbt]
\vskip-4.0cm
\centerline{{\vbox{\epsfxsize=16cm\epsfysize=12cm\epsfbox{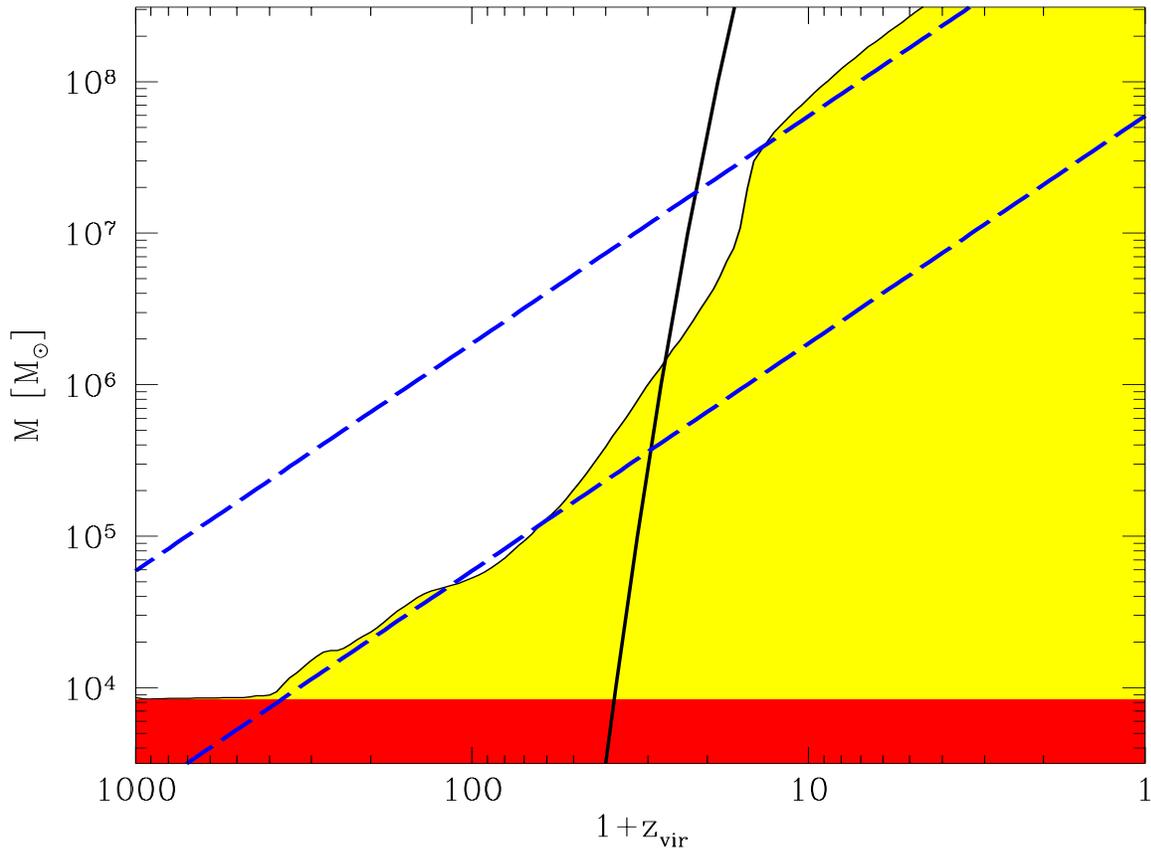}}}}
%\vskip-2cm
\caption{The minimum mass needed to collapse.}
The function $M_c(\zvir)$ is plotted as a function on
virialization redshift for standard CDM 
($\Omega=1$, $\Omega_b=0.06$, $h=0.5$).
%and BDM ($\Omega=\Omega_b=0.15$, $h=0.8$ -- light solid line). 
Only lumps 
whose parameters $(\zvir,M)$ lie above the shaded area can collapse and
form luminous objects. The dashed straight lines corresponding to $\Tvir=10^4\K$
and $\Tvir=10^3\K$ are shown for comparison (dashed). 
The dark-shaded region is that in which no radiative 
cooling mechanism whatsoever
could help collapse,
since $\Tvir$ would be lower than the CMB temperature.
The solid line corresponds to $3-\sigma$ peaks in standard CDM, normalized to 
$\sigma_8=0.7$, so such objects with baryonic mass 
%$\Omega_b\times 10^6\Ms \sim 6\tento{4}\Ms$ can form at $z\sim 30$.
$\Omega_b\times 2\times 10^6\Ms \sim 10^5\Ms$ can form at $z=30$.
\label{zmFig}
\end{figure}

The results of two sample runs are shown in 
figures\fignum{zfFig1} and\fignum{zfFig2}.
Both have $\Tvir=1000\K$ and the
standard CDM parameters $\Omega=1$, $\Omega_b=0.06$,
$h=0.5$. In \fig{zfFig1} (with $\zvir=100$), collapse
succeeds by our criterion below.
To the left, we see how recombination reduces the ionization 
fraction $x$ sharply at $z\sim 10^3$.
This weakens the Compton coupling to the CMB, and at 
$z\sim 400$, the gas temperature begins
dipping slightly below the CMB temperature (straight line).
At $z\sim 300$, a minute fraction of molecular hydrogen 
is formed via the $H_2^+$ channel before this reaction freezes out.
At $z\sim 100$, density and temperature rise to their virial values.
This causes a surge in the production of $H_2$ via the 
$H^-$ channel, producing a molecular abundance close to 
$10^{-3}$, and this in turn causes rapid cooling. From 
this point on, the curves in the figure 
are of course irrelevant, as the density will rise, causing even more
rapid cooling and a density profile that must ultimately be modeled with
a full 3D hydrodynamics simulation.

The evolution of a less successful lump is shown in 
\fig{zfFig2}, with $\zvir=10$.
Here even the molecules produced by the $H^-$-channel
around $z\sim 100$ are too few to cause significant
cooling. 
The molecules produced in the third wave wave of formation, 
at $z\sim\zvir$, are unable to cool the cloud substantially
simply because the density (and thus the cooling rate) has
become too low.
\fig{zfFig2}.

\subsection{The collapse criterion}
\def\dropfac{\eta}

We now give our operational definition of failure to collapse.
After the lump has virialized, we keep the density constant at 
$\rhovir$ and continue to integrate the equations for
the time evolution of temperature, ionization 
fraction and molecule abundance.
Loosely speaking, we consider the cloud a failure if 
its temperature has not dropped substantially within a Hubble 
time, which roughly corresponds 
to the redshift dropping by a factor $2^{2/3}$. 
We define failure to mean that 
\beq{CollapseDefEq}
T(\dropfac z) \ge \dropfac T(z),
\eeq
and choose $\dropfac=0.75$. 
We do not want to choose $\eta$ too small, since then even 
clouds that merely ``loiter" for a while and suddenly cool at
a substantially lower redshift (when molecule formation
suddenly becomes effective) will be counted as successful.

It should be noted that Compton cooling alone is useless for making
early structures. If it is able to cool the cloud substantially, 
the resulting contraction will drive up the recombination rate
(since the CMB temperature is $ << 10^4\K$),
virtually all free electrons will disappear, and Compton cooling 
will cease. Thus Compton cooling is self-destructive.
Molecular cooling does of course not suffer from this problem
once the $H_2$ has been produced, and can make runaway 
contraction proceed over many orders of magnitude.
The same goes for hydrogen line cooling: although it requires 
free electrons, the latter will be produced collisionally at
the high temperatures $\sim 10^4\K$ where line cooling is effective.

To ensure that our minimum mass is that above which 
runaway collapse (and thus possible formation of 
luminous objects) can occur, we thus ignore Compton cooling 
when $z<\zvir$.

\subsection{The ``shooting" scheme}

For each virialization redshift $\zvir$, there will 
clearly be some critical temperature $\Tvir$ such that clouds
with $T>\Tvir$ will collapse and clouds with 
$T<\Tvir$ will fail. 
We find this critical value by a ``shooting" scheme:
we run the code for a very high and a very low virial 
temperature, then again for the average of the two temperatures,
then use the interval halving method
to recursively home on on the critical value $\Tvir$. 
This is quite feasible numerically, since each individual
evolution run takes merely a few seconds on a workstation.

\section{DISCUSSION}

Earlier work on $H_2$ formation in the early universe has 
focussed on photodissociation  and subsequent suppression 
of $H_2$ cooling near the first structures to form, which 
are a likely source of ionizing photons at high redshift
(Silk 1985; Efstathiou 1992). 
Conversely, Haiman, Rees \& Loeb (1996) find that at low
redshift the ionizing background radiation field from
the first collapsing systems 
may actually stimulate $H_2$ formation and cooling in primordial clouds.
We have shown that 
% there is no need to posit such a source of photons in order 
even without ionizing radiation, 
enough electrons survive from the recombination epoch,
even in overdense collapsing regions,
for $H_2$ formation and cooling to be significant. 
Indeed, we have found that in the context of a standard CDM model, 
$H_2$ formation triggers cooling in virialized clouds and allows 
early formation of low mass objects.

Typical initial conditions for the first bound objects to form 
(from 3-$\sigma$ peaks) at $z\sim 30$ are found to be $f_{H_2}\sim 10^{-3}$,
$n_H\sim 10^2\cm^{-3}.$ Clouds of baryonic mass 
$\sim10^5 \Ms$ can be virialized at this redshift, with ensuing 
runaway $H_2$ cooling. The abundance of such objects 
is readily estimated by combining the Press-Schechter formalism
with the accurate derivation of the small-scale transfer function 
given by Hu \& Sugiyama (1995).
The impact of these first objects depends strongly
on unknown quantities such as their 
star formation initial mass function (IMF).
The most important issue is whether they were able to emit enough
ionizing radiation to reionize the universe or not. 
Below we will argue that they might have 
left observable imprints in both cases.

\subsection{If UV-emission is substantial...}

Let us first consider the former case, where UV-emission is substantial.
Our  population of condensed baryonic clouds could either undergo 
star formation or form massive black holes.
If the former fate awaits these clouds, it is plausible 
to believe, by analogy with our knowledge of the most 
metal-poor Galactic stars,
that a wide range of stellar masses is generated.
In either case, a substantial production of ionizing 
photons is likely.
In the former case, heavy elements will also be synthesized.
This would give a possible source for the 
heavy elements found at $z=2-4$
in Lyman alpha forest clouds, the most primitive objects 
in the universe, that amount to
$\sim 0.3\%$ of the solar abundances. 
% a fractional abundance $\sim 0.3\%\Omega_{Ly\alpha}.$
The IGM will be reheated at $z\simgt 10,$
thereby suppressing the formation of dwarf galaxies until 
a much later epoch, as argued by Blanchard {\etal} (1992). 
The low luminosity tail of the luminosity function of faint 
blue galaxies is indeed inferred to steepen with lookback time, 
as interpreted in models of faint galaxy number counts 
(Treyer and Silk 1994), consistent 
with recent ($z\sim 1$) formation.

In addition, optical depths of at least a few percent (Tegmark and Silk 1995) 
to electron scattering in the IGM are inevitable if reionization 
occurs when the first generation of objects condenses. 
This would lead to noteworthy implications for satellite 
proposals to measure the CMB anisotropy $C_l's$ to a precision 
of a percent or so. Scattering at this level would reduce the 
height of the acoustic peaks, which in the absence 
of early reionization are primarily 
sensitive to the baryon density.

\subsection{...and if it is not}

Let us now consider the latter case, where the initial UV-emission is negligible.
Even if star-formation is successful, there are at least three possible
things that could prevent substantial UV-emission: 
\begin{enumerate}
\item The IMF could be so steep that almost no OB-stars are formed.
\item The bulk of the UV-radiation could be absorbed locally, so that 
most of the radiation leaving the cloud is degraded below
the Lyman limit.
\item Since the clump would be quite loosely bound, with a virial temperature 
$\ll 10^4\K$, the first few massive stars might photoionize the entire cloud,
blow out the gas and thus prevent the bulk of the baryons from forming stars.
\end{enumerate}
If any of these caveats apply, then a much larger fraction ({\ie}, not just
$3-\sigma$ peaks) of the baryons would have time to form stars before 
global reionization finally raised the Jeans mass to above $10^4\K$ and terminated
this production of small objects. This turn-off might not occur 
until $z\sim 5-10$, which could leave as much as $50\%$ of the baryons in 
condensed MACHO-like objects. For a low density CDM cosmology 
with $\Omega\sim 0.3$ and a nucleosynthesis-favored baryon fraction
$\Omega_b\sim 0.06$, this would imply that 
about $10\%$ of our Galactic halo would consist of MACHOs.

\bigskip
\noindent
The authors would like to thank Martin Haehnelt, 
Uffe Hellsten, Avi Loeb
and Ned Wright for useful comments.
% And Angelica for endless patience with the M-O-L-E-C-U-L-E-S...
This work has been partially supported by European Union contract
CHRX-CT93-0120 and Deutsche Forschungsgemeinschaft
grant SFB-375.

\clearpage
\setcounter{secnumdepth}{-1}
\section{APPENDIX A}

In this appendix, we provide fits to the CMB photodissociation rates
of $H^-$ and $H_2^+$. 

The cross section for photodissociation of $H^-$ of Wishart (1979)
is well fit by the expression % (Shapiro \& Kang 1987)
\beq{HminusSigmaEq}
\sigma \aet{3.486}{-16}\cm^2\>\times {(x-1)^{3/2}\over x^{3.11}},
%\sigma \aet{7.928}{5}\cm^2\>\times {(\nu-\nu_0)^{3/2}\over \nu^3},
\eeq
where $x\equiv h\nu/0.74\eV$.

To accurately compute the photodissociation rate of $H_2^+$, one would
have to include photodissociation from all its excited states.  
However, due to lack of reliable molecular data, we only use the 
rates
for photodissociation from the ground state computed by Stancil
(1994), which we find to be well fit by the expression
\beq{H2plusSigmaEq}
\sigma \aet{7.401}{-18}\cm^2\>\times 10^{-x^2 - 0.0302 x^3 - 0.0158 x^4},
\eeq
where $x\equiv 2.762\ln(h\nu/11.05\eV)$. The cross section vanishes below the binding 
energy $h\nu=2.65$ eV. (Neglecting dissociation from excited
states will lead to a slight overestimate of the 
$H_2$-production though the $H_2^+$ channel.)
To obtain the desired dissociation rates $k$, we simply integrate the above cross sections
against a Planck spectrum:
\beq{PlanckIntegralEq}
k = {8\pi\over c^2} \int_0^\infty {\nu^2\sigma(\nu)d\nu\over e^{h\nu/k\Tg}-1},
\eeq
and fit the numerical results by the simple expressions given 
in Table 1.
$\badrate$ is accurate to within $10\%$ for the redshift range $40<z<2000$,
and $\badratetwo$ is correct to within $50\%$ for $150<z<1500$.

\clearpage
\section{REFERENCES}
 
{\frenchspacing

\rn Abel, T. 1995, Thesis, University of Regensburg.
% "Modelling Primordial Gas in Numerical Cosmology"

\rn
Abel, T., Anninos, P., Zhang, Y. \& Norman, M. L. 1996, submitted to New Astronomy.

\noindent
Anninos, P., Zhang, Y., Abel, T. \& Norman, M. L.
1996, to be submitted to New Astronomy.

\rf Bardeen, J. M., Bond, J. R., Kaiser, N. \& Szalay, A. S.
1986;ApJ;304;15
 
\rf Binney, J. 1977;ApJ;215;483

%\rf Black, J. 1981;MNRAS;197;553
 
\rf Blanchard, A., Valls-Gabaud, D. and 
Mamon, G. A. 1992;Astr. Ap.;264;365

\rf Bond, J. R. \& Szalay, A. S. 1983;ApJ;274;443

\rf de Araujo, J. C. N. \& Oph  er, R. 1988;MNRAS;231;923
% "Collapse of Population III objects"

\rf de Araujo, J. C. N. \& Opher, R. 1989;MNRAS;239;371
% "The masses of Population III objects"

\rf de Araujo, J. C. N. \& Opher, R. 1991;ApJ;379;461
% "Collapse of Population III objects induced by cold collisionless dark matter"
J.C.N. de Araujo \& R. Opher
ApJ (1991), 379, 461-465

\rf Efstathiou, G. 1992;MNRAS;256;43P

% \rf Cen, R., Gnedin, N. Y., Koffmann, L. A., \& Ostriker, J. P. 1992;ApJ;399;L11
% A tilted cold dark matter cosmological scenario.
% Astrophysical Journal, Letters, 1 Nov. 1992, vol.399, (no.1, pt.2):L11-14.
% Pub type:  Theoretical or Mathematical.

\rf Cen, R., Ostriker, J. P. \& Peebles, P. J. E 1993;ApJ;415;423
% Cen, R.; Ostriker, J.P.; Peebles, P.J.E.
% A hydrodynamic approach to cosmology: the primeval baryon isocurvature model.
% Astrophysical Journal, 1 Oct. 1993, vol.415, (no.2, pt.1):423-44.

\rf Cole, S., Aragon-Salamanca, A., Frenk, C.S., Navarro, J.F. \& Zepf, S.E. 1994;MNRAS;271;781
 
\rf Couchman, H. M. P. 1985;MNRAS;214;137

\rf Couchman, H. M. P. \& Rees, M. J. 1986;MNRAS;221;53

\rf Dalgarno, A. \& McCray, R. A. 1972;ARA\&A;10;375
 
\rf Fukugita, M \& Kawasaki M 1991;MNRAS;269;563
% EARLY REIONIZATION. 

\rf Gnedin, N. Y. \& Ostriker, J. P.1992;ApJ;400;1

\rf Gunn, J. E. and Peterson, B. A. 1965;ApJ;142;1633

\rf Haiman, Z., Thoul, A. A. \& Loeb, A. 1996;ApJ;464;523
% preprint astro-ph/9507111.
% HAIMAN Z; THOUL AA; LOEB A.
% COSMOLOGICAL FORMATION OF LOW-MASS OBJECTS.
% ASTROPHYSICAL JOURNAL, 1996 JUN 20, V464 N2:523-538.
     
\rf Haiman, Z., Rees, M. \& Loeb, A. 1996;ApJ;467;522
% preprint astro-ph/9511126.
% HAIMAN Z; REES MJ; LOEB A.
% H-2 COOLING OF PRIMORDIAL GAS TRIGGERED BY UV IRRADIATION.
% ASTROPHYSICAL JOURNAL, 1996 AUG 20, V467 N2:522-531.
     
\rf Hirasawa, T. 1969; Prog. Theor. Phys.;42;523

\rf Hollenbach, D. and McKee, C. F. 1979;ApJS;41;555

\rf Hu, W., Scott, D. \& Silk,  J. 1994;Phys. Rev. D;49;648
% REIONIZATION AND COSMIC MICROWAVE BACKGROUND DISTORTIONS - A COMPLETE
% TREATMENT OF SECOND-ORDER COMPTON SCATTERING.

\rn Hu, W. \& Sugiyama, N. 1995, preprint astro-ph/9510117

\rf Hutchins, J. B. 1976;ApJ;205;103

\rf Karpas, Z., Anicich, V. \& 
Huntress, W. T. Jr. 1979;J. Chem. Phys.;70;2877

\rf Kauffmann, G, White, S.D.M. \& Guiderdoni, B. 1993;MNRAS;264;201

\rf Lepp, S. \& Shull, J. M. 1983;ApJ;270;578

\rf Liddle, A. R. \& Lyth, D. H. 1995;MNRAS;273;1177
% astro-ph/9409077 
% Trends in large-scale structure observations and the likelihood of early
% reionization.
% Monthly Notices of the Royal Astronomical Society, 15 April 1995, vol. 273,
% (no.4):1177-84.
% (REFERS TO US AND FUKUGITA & KAWASAKI)

\rn Loeb, A. 1996, private communication.

\rf Mather {\etal} 1994;ApJ;420;439
% MEASUREMENT OF THE COSMIC MICROWAVE BACKGROUND SPECTRUM BY
% THE COBE FIRAS INSTRUMENT.

\rn Norman, M.L., Abel, T.,  Zhang, Y. \&  Anninos, P.
1996, to be submitted to New Astronomy.

\rf Puy, D., Alecian, G, {\etal} 1993;A\&A;267;337
% Title:             Formation of primordial molecules and thermal balance
%                    in the early universe
% Authors:           PUY, D.; ALECIAN, G.; LE BOURLOT, J.; LEORAT, J.;
%                    PINEAU DES FORETS, G.
% Journal:           Astronomy and Astrophysics (ISSN 0004-6361), vol. 267,
%                    no. 2, p. 337-346.
% Publication Date:  01/1993

\rf Puy, D. \& Signore, M. 1996;A\&A;305;371
% PUY D; SIGNORE M.
% PRIMORDIAL MOLECULES IN THE EARLY CLOUD FORMATION.
% ASTRONOMY AND ASTROPHYSICS, 1996 JAN 10, V305 N2:371-378.
     
% Primordial molecules in early cloud formation.
% Larger masses, different focus, terrible English,
% but extensive compilation of rates
%: Comment: the rates are also not that great ;-)

\rf Peebles, P. J. E. 1987; ApJ;315;L73

\rn Peebles, P. J. E. 1993, Principles of Physical Cosmology
(Princeton University Press)

\rf Press, W. H. \& Schechter, P. 1974;ApJ;187;425
 
\rf Rees, M. J. \& Ostriker, J. P. 1977;MNRAS;179;541
% The tcool=tgrav bible 
 
\rf Seaton, M. 1959;MNRAS;119;84

\rf Shapiro, P. R. \& Kang, H. 1987;ApJ;318;32

\rf Silk, J. 1977;ApJ;211;638
 
\rf Silk, J. 1983;MNRAS;205;705
% THE FIRST STARS (contains useful rates)

\rf Silk, J. 1985;ApJ;297;1

\rf Smoot, G. F. {\etal} 1992;ApJ;396;L1
% STRUCTURE IN THE COBE DMR 1ST YEAR MAPS

\rf Stancil, P. C. 1994;ApJ;430;360

\rf Steidel, C. C. and Sargent, W. L. W. 1987;ApJ;318;L11

\rf Subrahmanyan, R. {\etal} 1993;MNRAS;263;416
%     SUBRAHMANYAN R; EKERS RD; SINCLAIR M; SILK J.
%     A SEARCH FOR ARCMIN-SCALE ANISOTROPY IN THE COSMIC MICROWAVE BACKGROUND.

\rn Tegmark, M. 1994, Ph.D. Thesis, U.C. Berkeley

\rf Tegmark, M. \& Silk, J. 1994;ApJ;423;529
% y.tex

\rf Tegmark, M. \& Silk, J. 1995;ApJ;441;458
% openreion

\rn Tegmark, M., Silk, J. \& 
Blanchard, A. 1994, ApJ, {\bf 420}, 484; {\bf 434}, 395.
%\rf Tegmark, M., Silk, J. \& Blanchard, A. 1994;ApJ;420;484
%     ON THE INEVITABILITY OF REIONIZATION - IMPLICATIONS FOR COSMIC MICROWAVE
%     BACKGROUND FLUCTUATIONS.
%     ASTROPHYSICAL JOURNAL, 1994 JAN 10, V420 N2:484-496.

\rf Treyer, M.\& Silk, J. 1994;ApJ;436;L143

\rf Wishart, A. W. 1979;MNRAS;187;59P

\rf Vishniac, E. 1987;ApJ;322;597
 
\rf Webb, J. K., Barcons, X., Carswell, R. F., and
Parnell, H. C. 1992;MNRAS;255;319

\rf White, S. D. M. \& Rees, M. J. 1978;MNRAS;183;341
% Core condensation in heavy halos: a two-stage theory for galaxy formation
% and clustering.

% Perhaps these too:

%\rn Matsuhada, Sato \& Takeda 1971
% Recombination rates 

% \rf Stebbins, A., \& Silk, J. 1986;ApJ;300;1

}

\end{document}